\documentclass[12pt]{article}
\def\reptitle{SUPERSYMMETRIZATION SCHEMES OF MAXWELL ALGEBRA}

\usepackage{bbm}
\usepackage{amsmath}
\catcode`@=11
\def\secteqno{\@addtoreset{equation}{section}%
\def\theequation{\thesection.\arabic{equation}}}
\catcode`@=12
\secteqno
\newcommand{\be}{\begin{equation}}
\newcommand{\ee}{\end{equation}}
\newcommand{\bea}{\begin{eqnarray}}
\newcommand{\eea}{\end{eqnarray}}
\newcommand{\beann}{\begin{eqnarray*}}
\newcommand{\eeann}{\end{eqnarray*}}
\newcommand{\bi}{\begin{enumerate}}
\newcommand{\ei}{\end{enumerate}}
\newcommand{\bref}[1]{(\ref{#1})}
\newcommand{\nn}{\nonumber}
\newcommand{\A}{\alpha} \newcommand{\B}{\beta} \newcommand{\gam}{\gamma}
\newcommand{\G}{\Gamma} \newcommand{\D}{\delta} 
\newcommand{\ep}{\epsilon}

 \newcommand{\s}{\sigma}

\newcommand{\h}{\eta}           
           
\newcommand{\W}{\Omega}

\newcommand{\Gam}{\Gamma}

\newcommand{\ba}{\bar }
\def\6{\partial}\def\7{\tilde}\def\8{\hat}

\def\CS{{\cal S}}
\def\CT{{\cal T}}
\def\CB{{\cal B}}

\def\CP{{\cal P}}\def\CJ{{\cal J}}
\def\l{{\ell}}
\def\CM{{\cal M}}\def\={{\;=\;}}\def\CQ{{\cal Q}}

\def\vs{\vskip 4mm}\def\={{\;=\;}}\def\+{{\;+\;}}
\def\rB{{{B}}}\def\bJ{{\bf{J}}}\def\bB{{\bf{B}}}
\def\bQ{{\rm{\bf Q}}}\def\bSig{{\bf {\Sigma}}}
\def\bS{{\bf {S}}}\def\bT{{\rm{\bf T}}}

\def\SO{O}
\def\CBA{{{\cal T}_0}}
\def\CBB{{{\cal T}_5}}

\def\ebox#1#2{\vskip 2mm{\vbox{\hrule\hbox{\vrule\kern3pt\vbox{\kern3pt
         {\begin{eqnarray}#1\label{#2}\end{eqnarray}}
         \kern3pt}\kern3pt\vrule}\hrule}}\vskip 2mm}
\def\tbox{\vskip 2mm{\vbox{\hrule\hbox{\vrule\kern3pt\vbox{\kern3pt
         {{\hfill {\small ${}^{notebook\;}$} \\
         \large \bf ~~\reptitle}\\ } 
         \kern3pt}\kern3pt\vrule}\hrule}}\vskip 2mm}

\def\vs{\vskip 4mm}
\def\half{{\textstyle\frac{1}{2}}}
\begin{document}
{~~ }\vskip 1cm
\begin{center}
{\Large SUPERSYMMETRIZATION SCHEMES OF $D=4$ \\ \vskip 3mm MAXWELL ALGEBRA }
\vskip 8mm
{\large {Kiyoshi Kamimura${}^1$} and {Jerzy Lukierski${}^2$}}\vskip 8mm
{\it
${}^1$  Department of Physics, Toho University Funabashi274-8510, Japan \\
${}^2$Institute of Theoretical Physics, Wroclaw University, pl.
Maxa Borna 9, 50-204 Wroclaw, Poland}
\vskip 5mm
{\rm  kamimura@ph.sci.toho-u.ac.jp,  lukier@ift.uni.wroc.pl}
\end{center}

\begin{abstract}
{The Maxwell algebra, an  enlargement of Poincar\'{e} algebra by Abelian tensorial generators, can be obtained in arbitrary dimension $D$ by the suitable contraction of $O(D-1,1) \oplus O(D-1,2)$ (Lorentz algebra$\oplus AdS$ algebra).
We recall that in $D=4$ the Lorentz algebra $O(3,1)$ is described by the realification $Sp_R(2|C)$ of complex algebra $Sp(2|C)\simeq {Sl(2|C)}$ and $O(3,2)\simeq Sp(4)$.
We study various $D=4$ $N$-extended Maxwell superalgebras obtained by the contractions of real superalgebras ${OSp}_R(2N-k; 2|C)\oplus{OSp}(k;4)$, ($k=1,2,\ldots,2N$) (extended Lorentz superalgebra $\oplus$ extended AdS superalgebra).

 If $N=1$ ($k=1,2$) one arrives at two different versions of simple Maxwell superalgebra. For any fixed $N$ we get $2N$ different superextensions of Maxwell algebra with $n$-extended Poincar\'{e} superalgebras ($1\leq n \leq N$) and the internal symmetry sectors obtained by suitable contractions of the real algebra $O_R(2N-k|C)\oplus O(k)$. Finally the comments on possible applications of Maxwell superalgebras are presented.}
\end{abstract}

\section{Introduction}

The Maxwell algebra as the enlargement of Poincar\'{e} algebra ($P_\mu, M_{\mu\nu}$) by antisymmetric Abelian tensorial charges $Z_{\mu\nu} =-Z_{\nu\mu}$ was firstly obtained for $D=4$ more than forty years ago \cite{Bacry:1970ye,Schrader:1972zd} by supplementing the relations
\begin{eqnarray}\label{wroja1.1}
[P_\mu, P_\nu ]  &= &  i\, M^2 \, Z_{\mu\nu}\, ,
\label{wroja1a}
\nn\\
 \left[ P_{\rho}, Z_{\mu\nu} \right] & = & 0, \qquad  \qquad
 [ M_{\rho\tau} , Z_{\mu\nu} ] =-i(\h_{\mu[\tau}\,Z_{\rho]\nu}-
\h_{\nu[\tau}\,Z_{\rho]\mu})\,,
\label{wroja1b}
\end{eqnarray}
where $M$ is a geometric mass parameter ($[M]=1;$ we use in $D=4$ the metric $\eta_{\mu\nu}={\rm diag}(-1,1,1,1))$.
In  presenting the  Maxwell algebra we introduce mass-like fundamental parameter $M$ (in \cite{biwroja3} denoted by $\Lambda=M^2$) which implies vanishing mass dimensionality of the tensorial charges $([Z_{\mu\nu}]=0)$. The relations (\ref{wroja1a}) can be however rewritten for generators $Z_{\mu\nu}$ with any mass dimensionality. In particular if $M^2$ is replaced by electromagnetic dimensionless coupling constant $e$ ($[e]=0$; see e.g. \cite{Bacry:1970ye,Schrader:1972zd,biwroja4new,Gomis:2009vm}) we obtain $[Z_{\mu\nu}]=2$. Simple dynamical realization of such Maxwell algebra  is obtained in the model of relativistic free particle moving in constant EM field backgrounds \cite{biwroja4new,Gomis:2009vm}

We recall that Maxwell algebra can be obtained by a contraction of $O(3,1)\oplus O(3,2)$ \cite{Soroka:2006aj,Gomis:2009vm}\footnote{Recently deformation $O(3,1)\oplus O(3,2)$ of Maxwell algebra was called $AdS$-Maxwell algebra \protect\cite{biwroja7}.} or $O(3,1)\oplus O(4,1)$   \cite{Gomis:2009vm} (if we replace $M^2 \to - M^2$ in \bref{wroja1.1}).
  Recently in \cite{Bonanos:2009wy} there were as well introduced the simple Maxwell superalgebras, with two Weyl supercharges ${\bf Q}_\alpha , {\bf \Sigma}_\alpha$, where ${\bf Q}_\alpha$ are the standard $N={1}$ Poincar\'{e} supercharges and new Maxwell supercharges  ${\bf \Sigma}_\alpha$ are required for the supersymmetrization of the generators  $Z_{\mu\nu}$. One should point out that recently appeared proposals to use the Maxwell algebra \cite{biwroja3,biwroja9,biwroja7,Durka:2011va} and Maxwell superalgebra \cite{biwroja10} to the geometric extension of (super)gravity  theories.

  At present it appears interesting to study the  problem how the Maxwell algebra can be supersymmetrized in various ways. In order to obtain important class of $D=4$ Maxwell superalgebras we observe that $D=4$ Lorentz algebra can be obtained as the realification\footnote{
By realification of n-dimensional complex Lie (super)algebra $\widehat{g}$ with the generators $g_i=g^{(1)}_i + i g^{(2)}_i, (i=1,...,n)$ we call the  2n-dimensional real Lie (super)algebra $\widehat{g}_R$ {on real space} with real generators $g^{(1)}_i, g^{(2)}_i$ (see e.g. \cite{Onishchik:1991}).
 If besides the complex generators $g_i$ we introduce $\ba g_i=g^{(1)}_i - i g^{(2)}_i$, the real  generators of $\8g_R$ are linear combinations of
 $g_i$ and $\ba g_i$, ($g^{(1)}_i=\frac12(g_i+\8g_i),\;g^{(2)}_i=\frac1{2i}(g_i-\8g_i)).$}
of complex algebra $Sp(2|C)$ ($O(3,1)\simeq Sp_R(2|C)$) with the supersymmetrization described by the realification of complex superalgebra $OSp(m;2|C)$,
(for $m=1$ see e.g. \cite{biwroja11}); further it is well known that $D=4$ $AdS$ symmetries are supersymmetrized by $D=4$ $AdS$ superalgebras $OSp(k;4)$.
   We shall consider therefore contractions\footnote{
We need to consider the contractions of direct sum $\hat{g}_1 \oplus \hat{g}_2$ of  Lie (super)algebras in which the contracted generators appear as suitable sums of the generators  belonging to $\hat{g}_1$  and $\hat{g}_2$.
For earlier discussion of contractions which use the linear superposition of generators from the sum of Lie (super)algebras, see \cite{biwroja12,biwroja13,biwroja14}.
}
of the following choices of the sum of real semisimple superalgebras
\begin{equation}
\label{wroja1.3} O(3,1)\oplus O(3,2)\quad \longrightarrow^{\hskip -9mm {SUSY}}\quad  OSp_R(2N-k; 2|C) \oplus  OSp(k;4)
\end{equation}
where $k=0,1,2\ldots 2N$ and  $N=0,1,2\ldots$.

We assume that the derived extended Maxwell superalgebras should satisfy the following two properties:
\begin{description}
\item{i)} They should contain at least one standard bilinear fermionic SUSY relation characterizing Poincar\'e supercharges with the following algebraic structure
    \begin{equation}
    \label{wroja1.4}
\left\{ \bQ_{\alpha}, {\bQ}_{{\beta}} \right\}
    = (C\gamma^{\mu} \, P_{\mu})_{\alpha {\beta}}+{\rm\;central\; charges}.
    \end{equation}
        \item{ii)} Their bosonic sector should contain the Maxwell algebra.
    \end{description}
        Let us observe that in the contraction of \bref{wroja1.3}
    the Poincar\'{e} supercharges satisfying the relations \bref{wroja1.4}
    can be only obtained from the contraction of $OSp(k;4)$.
    If we put $k=0$ in \bref{wroja1.3}
the generators $Sp(4)\simeq O(3,2)$ remain not supersymmetrized and we can get by contraction only a purely exotic version of Maxwell superalgebra \cite{biwroja15,biwroja13} with the  standard generic SUSY relations \bref{wroja1.4} replaced by the following one

\begin{equation}
\label{wroja1.5}
\{ {\bf S}_{\alpha}, {\bf S}_{\beta} \}
= (C\gamma^{\mu\nu} Z_{\mu\nu})_{\alpha\beta}\,+{\rm\;central\; charges}.
\end{equation}

The paper is organized as follows: In Sect.~2 we shall derive the simple $N=1$  Maxwell superalgebras. If we put $N=1$ in \bref{wroja1.3}, for $k=2$ and $k=1$ we shall obtain by contractions two versions of simple ($N=1$) $D=4$ Maxwell superalgebras, differing by the presence of Abelian chiral generator (see also \cite{Bonanos:2009wy,biwroja14}).
   In Sect.~3 we consider the nonstandard contraction of \bref{wroja1.3} in general case (arbitrary $N$ with $k=1,2,\ldots,2N$); we treat
 separately the cases $0\leq k< N$ and $N\leq k\leq 2N$ requiring different explicit  contractions.   In Sect~4 we provide short discussion of our results.

\section{$N=1$ Maxwell superalgebras by contraction}

\subsection{Simple Maxwell superalgebras}

In this section we consider the contractions of the superalgebras \bref{wroja1.3}
for $N=1$ and $k=1,2$ providing the simple Maxwell superalgebras with two Weyl supercharges which were proposed in
\cite{Bonanos:2009wy}. The Maxwell algebra (relations (\ref{wroja1a})+Lorentz algebra) is extended supersymmetrically as follows\footnote{
The mass dimension of the generators are $[P_{\mu}]=1,\;[Z_{\mu\nu}]=0,\; [\bQ_\A]=\frac12,\;   [\bS_\A]=\frac32,\; [B]=2,\; [B_5]=0$.}
\bea
\{{\bQ}_\A,{\bQ}_\B\}&=&(C\gamma^{\mu})_{\A\B}P_\mu,\qquad
\{{\bSig}_\A,{\bSig}_\B\}=0,
\nn\\
\{{\bQ}_\A,{\bf\Sigma}_{\B}\}&=&\frac{M^2}2(C\gamma^{\mu\nu})_{\A\B}\,Z_{\mu\nu}\,+\, (C\gamma_5)_{\A\B}\,{\rB},
\label{Maxsuperalgebra-1}
\\
\left[P_\mu,{\bQ}_\A\right]&=&-\frac{i}2\,\bSig_{\B}(\gamma_{\mu}{)^{\B}}_\A,\qquad
\left[P_\mu,{\bSig}_\A\right]=0,
\nn\\
\left[M_{\rho\s},\bQ_{\A}\right]&=&-\frac{i}2(\bQ \gamma_{\rho\s})_{\A},\qquad
\left[M_{\rho\s},\bSig_{\A}\right]=-\frac{i}2(\bSig \gamma_{\rho\s})_{\A}\,,
\label{Maxsuperalgebra-2}
\\
\left[B_C,{\bQ}_\A\right]&=&i\,(\bQ\gamma_{5})_\A,\qquad
\left[B_C,{\bSig}_\A\right]=-\,i\,(\bSig\gamma_{5})_\A\,,
\label{Maxsuperalgebra}
\eea
where real Dirac-Majorana matrices $\gamma_{\mu}$
 ($\gamma^T_i=\gamma_i, \gamma^T_0=-\gamma_0$) verify
 the $O(3,1)$ Clifford algebra $\{ \gamma_\mu, \gamma_\nu \}=2\eta_{\mu\nu}$ and $\gamma_{\mu\nu}=\half [ \gamma_\mu, \gamma_\nu ]= -\gamma_{\nu\mu}$ is the $4\times 4$ matrix realization of $O(3,1)$.
 The charge conjugation matrix $C=\gamma_0$ and $\gamma_5 = \gamma_0 \gamma_1 \gamma_2 \gamma_3 = - \gamma^T_5$ satisfy the relations
$(C\gamma_\mu)^T=(C\gamma_\mu),\;(C\gamma_{\mu\nu})^T=(C\gamma_{\mu\nu}),\;
(C\gamma_5)^T=-(C\gamma_5).
$ The new supercharges ${\bf \Sigma}_\alpha$ are  needed for the supersymmetrization of the generators   $Z_{\mu\nu};$ the $B_C$ describes a chiral generator and $B$ is the central charge. In the minimal $N=1$ Maxwell superalgebra we put $B=B_C=0$.

\subsection{Contraction of $O(3,1) \oplus  OSp(2;4)\;\;(k=2$ case)}

The $OSp(2,4)$ superalgebra is
\bea
\left[\CM_{{\8\mu}{\8\nu}},\CM_{\8\rho\8\s}\right]&=&
-i\,\h_{\8\rho[{\8\nu}}\CM_{{\8\mu}]\8\s}+i\,
\h_{{\8\s}[\8\nu}\CM_{{\8\mu}]\8\rho},
\label{OSp04}\\
\{{\CQ}_{\A}^{ i},{\CQ}_{\B}^{ j}\}&=&-\,\D^{ij}\,(C\Gam^{{\8\mu}{\8\nu}})_{\A\B}
\CM_{{\8\mu}{\8\nu}}+\,2\,(C)_{\A\B}\CB^{ij},
\nn\\
\left[\CM_{{\8\mu}{\8\nu}},\CQ_{\A}^{ i}\right]&=&-\frac{i}2(\CQ^i \Gam_{{\8\mu}{\8\nu}})_{\A},
\qquad
\left[\CB^{ij},{\CQ}_{\A}^{k}\right]=-i\,\D^{k[j}\,\CQ_{\A}^{ i]}.
\label{OSp24}\eea
Here $ \CM_{{\8\mu}{\8\nu}},\, (\8\mu=0,1,2,3,4) $ are $SO(3,2)$ generators with  $\h_{{\8\mu}{\8\nu}}=(-1,1,1,1,-1)$ and
$O(3,1)$ real Dirac-Majorana matrices
$\Gamma_\mu = \gamma_{\mu}\gamma_{5}$, $\Gamma_4= -\gamma_5$
 satisfy the $O(3,2)$ Clifford algebra
 $\{\Gamma_{\hat{\mu}}, \Gamma_{\hat{\nu}} \}= 2\eta_{\hat{\mu} \hat{\nu}}$.
 The $4\times 4$ matrix realization $\Gamma_{\hat{\mu}\hat{\nu}}=\half[ \Gamma_{\hat{\mu}}, \Gamma_{\hat{\nu}} ]$ of $O(3,2)$ algebra is expressed by
 $O(3,1)$ $\gamma_\mu$-matrices $ as
 \Gamma_{\mu\nu} = \half [ \Gamma_\mu , \Gamma_\nu ]
 = \gamma_{\mu\nu}, \;
 \Gamma_{\mu4}= \Gamma_\mu \Gamma_4 = \gamma_\mu.$
The charge conjugation $C=\gamma_0 = \Gamma_0 \Gamma_4$ is common
 for $O(3,2)$ and $O(3,1)$.
Supercharges ${\CQ}_{\A}^{i},\,(\A=1,2,3,4,\,i=1,2)$ are real $O(3,2)$ spinors and $\CB^{ij}=-\ep^{ij}\CB_C$ is the $SO(2)$ generator.

 The real algebra $O(3,1)$ is provided by
the algebra $Sp_R(2|C)=O_R(2,1|C) \simeq O(3,1)$.
The algebra $O(2,1|C)\oplus{\overline{O(2,1|C)}}$ is described by the generators $\bJ_{\ba\mu}$ and $\bJ^\dagger_{\ba\mu}, ({\ba\mu}=0,1,2)$ with the metric $\h_{{\bar\mu}{\bar\nu}}=(-1,1,1)$,
\bea
\left[\bJ_{\ba\mu},\bJ_{\ba\nu}\right]={i}\,{\ep_{\ba\mu\ba\nu}}^{\ba\rho}
\bJ_{\ba\rho},\qquad \left[\bJ^\dagger_{\ba\mu},\bJ^\dagger_{\ba\nu}\right]={i}\,{\ep_{\ba\mu\ba\nu}}^{\ba\rho}\bJ^\dagger_{\ba\rho},\qquad
[\bJ_{\ba\mu},\bJ^\dagger_{\ba\nu}]=0,
\label{OSp02}\eea
where  $\ep^{\ba\mu\ba\nu\ba\rho}$ is the Levi-Civita symbol with $\ep^{012}=-\ep_{012}=1$.
If we introduce the real $O(3,1)$ generators
$\CJ_{\mu\nu}, \,(\mu=0,1,2,3)$ by the relation
\be
\bJ_{\ba\mu}=\frac14{\ep_{\ba\mu}}^{\ba\rho\ba\nu}
J_{\ba\rho\ba\nu}+\frac{i}2\,J_{\ba\mu 3}\,,
\label{bJdef}
\ee
they satisfy the $D=4$ Lorentz algebra  with $\h_{\mu\nu}=(-1,1,1,1)$,
\bea
\left[\CJ_{{\mu}{\nu}},\CJ_{\rho\s}\right]&=&-i\,
\h_{{\rho}[\nu}\CJ_{{\mu]}\s}+i\,\h_{{\s}[\nu}\CJ_{{\mu]}\rho}\,.
\label{O31}\eea
We propose the following redefinitions of the generators of $O(3,1)\oplus O(3,2)$ (we recall that $\Gamma_4=-\gamma_5$ and
we put $\A+\B=1$) \be
M_{\mu\nu}= \CJ_{\mu\nu} + \CM_{\mu\nu},\quad
P_\mu  = \frac1R\,{\CM_{\mu 4}},\quad
Z_{\mu\nu} =  \frac{1}{R^2 M^2} (\A\CJ_{\mu\nu}-\B{\CM}_{\mu\nu})
\label{relgenSM3Rplus-1}
\ee
and the rescaled internal symmetry generator and supercharges as follows
\bea
B_C&=&\frac1{R^{\gamma}} {\CB}_C\,, \label{relgenSM3RplusB}
\\
 {\bf Q}_{\alpha} = \frac{1}{2R^{1/2}} (Q^1_\alpha+( Q^2\gamma_5 )_{\alpha}),
&\,& d
{\bf \Sigma}_{\alpha} = \frac{1}{2R^{3/2}} (Q^1_\alpha-(Q^2\gamma_5)_{\alpha}) \,,\label{relgenSM3Rplus}
\eea
where $R$ in the formulas \bref{relgenSM3Rplus-1}-\bref{relgenSM3Rplus}
is a contraction parameter with dimension of length $([R]={-1})$. The superalgebra $O(3,1)\oplus OSp(2;4)$ in terms of new rescaled real generators takes the form
\bea
\left[P_\mu,P_{\nu}\right]&=&i\,M^2\,Z_{\mu\nu}-i\frac{\B}{R^2}\,M_{\mu\nu},\qquad
\left[P_\mu,M_{\rho\s}\right]=-i\,\h_{\mu[\rho}P_{\s]},\nn\\
\left[P_\mu,Z_{\rho\s}\right]&=&i\,\frac{\A}{R^2}\,\h_{\mu[\rho}P_{\s]},\qquad
\left[M_{\mu\nu},Z_{\rho\s}\right]=-i\,\h_{\rho[\nu}Z_{\mu] \s}+i\,
\h_{\s[\nu}Z_{\mu]\rho},\nn\\
\left[M_{\mu\nu},M_{\rho\s}\right]&=&-i\,\h_{\rho[\nu}M_{\mu] \s}+i\,
\h_{\s[\nu}M_{\mu]\rho},\nn\\
\left[Z_{\mu\nu},Z_{\rho\s}\right]&=&-i\frac{\B-\A}{M^2\,R^2}
(\,\h_{\rho[\nu}Z_{\mu] \s}-\,\h_{\s[\nu}Z_{\mu]\rho})
-i\frac{\B\A}{M^4\,R^4}(\h_{\rho[\nu}M_{\mu] \s}-\,
\h_{\s[\nu}M_{\mu]\rho}),\label{BMaxalgeb02}
\eea\bea
\{{\bQ}_\A,{\bQ}_\B\}&=&(C\gamma^{\mu})_{\A\B}P_\mu,\qquad
\{{\bSig}_\A,{\bSig}_\B\}=\frac1{R^2}\,(C\gamma^{\mu})_{\A\B}P_\mu,
\nn\\
\{{\bQ}_\A,{\bf\Sigma}_{\B}\}&=&\frac{M^2}2(C\gamma^{\mu\nu})_{\A\B}\,Z_{\mu\nu}\,+\,\frac1{R^{2-\gam}}\, (C\gamma_5)_{\A\B}\,B_C,
\\
\left[P_\mu,{\bQ}_\A\right]&=&-\frac{i}2\,\bSig_{\B}(\gamma_{\mu}{)^{\B}}_\A,\qquad
\left[P_\mu,{\bSig}_\A\right]=-\frac{i}{4R^2}\,\bQ_{\B}(\gamma_{\mu}{)^{\B}}_\A,
\nn\\
\left[Z_{\mu\nu},{\bQ}_\A\right]&=&\frac{i}{2R^2 M^2}\,(\bQ\gamma_{\mu\nu})_\A,\qquad
\left[Z_{\mu\nu},{\bSig}_\A\right]=\frac{i}{2R^2 M^2}\,(\bSig\G_{\mu\nu})_\A,
\nn\\
\left[M_{\rho\s},\bQ_{\A}\right]&=&-\frac{i}2(\bQ \gamma_{\rho\s})_{\A},\qquad
\left[M_{\rho\s},\bSig_{\A}\right]=-\frac{i}2(\bSig \gamma_{\rho\s})_{\A}.
\nn\\
\left[B_C,{\bQ}_\A\right]&=&\frac{i}{R^{\gam}}\,(\bQ\gamma_{5})_\A,\qquad
\left[B_C,{\bSig}_\A\right]=-
\, \frac{i}{R^{\gam }}\,(\bSig\gamma_{5})_\A \,.
\label{kalgebra}
\eea
If we put $\A=1, \B=0, $ \,
 $2>\gam >0$  and $M=1$  the formulas \bref{BMaxalgeb02}-\bref{kalgebra} describe  the k-deformation
of the Maxwell superalgebra
given in \cite{biwroja14}($\frac{1}{R^2}=$k), which reproduces in the k$\to 0$ limit the Maxwell superalgebra (\ref{Maxsuperalgebra-1}-\ref{Maxsuperalgebra}) with $B_C=0$ and  $B=0$.
In the case $\gamma=0$, $B_C$ is nonvanishing and becomes a chiral charge generating
chiral transformation of $\bQ_\A$ and $\bSig_\A$, (see \bref{Maxsuperalgebra}).
The specific feature of our contraction is that two factors from the defining algebra $OSp(2|4)\oplus \SO(3,1)$ are suitably mixed  (see \bref{relgenSM3Rplus-1}--\bref{relgenSM3Rplus}).
As a result, this type of contractions will not respect the direct sum structure of the uncontracted algebras
(see also \cite{Cangemi:1992ri},\cite{deAzcarraga:2002xi} (Sect.~8)).


\subsection{Contraction of $OSp_R(1; 2|C) \oplus  OSp(1;4)\;\; (k=1$ case) }

The  $OSp(1;4)$ superalgebra is
\bea
\left[\CM_{{\8\mu}{\8\nu}},\CM_{\8\rho\8\s}\right]&=&
-i\,\h_{{\8\rho}[\8\nu}\CM_{{\8\mu}]\8\s}+i\,
\h_{{\8\s}[\8\nu}\CM_{{\8\mu}]\8\rho},
\nn\\
\{{\CQ}_{\A}^{},{\CQ}_{\B}^{}\}&=&-\frac{1}{2}\,(C\Gam^{{\8\mu}{\8\nu}})_{\A\B}
\CM_{{\8\mu}{\8\nu}},\qquad
\left[\CM_{{\8\mu}{\8\nu}},\CQ_{\A}\right]=-\frac{i}2(\CQ\Gam_{{\8\mu}{\8\nu}})_{\A}.
\eea
The complex superalgebra $OSp(1;2|C)=(\bJ_{\ba\mu},{ S}_{+\A})$  contains the bosonic subalgebra
$Sp(2|C)=O(2,1|C)$ with its realification
given by $O(3,1)$. The complex supercharges ${ S}_{+\A}$ define the
fermionic sector of $OSp(1;2|C)$,
\bea
\{ { S}_{+\A},{ S}_{+\B}\}
&=&4i\,(C\G_+^{\ba\rho})_{\A\B}\,\bJ_{\ba\rho},
\qquad
\left[\bJ_{\ba\mu},{ S}_{+\A}\right]=\frac12\,{{
 S}_{+\B}\,({\Gamma_{+\ba\mu}})^\B}_\A,
\label{OSp12C}\eea
where the matrices ${\Gamma_{+\ba\mu}}$ are defined by O(3,1) gamma matrices $\gamma_\mu$ as
\bea
{\G}_{+\ba\mu}&=&\gamma_{\ba\mu}\gamma_3\,P_+,\quad P_\pm=\frac12(1\pm i\,\gamma_5),\quad
[{\G}_{+\ba\mu},{\G}_{+\ba\nu}]=2i\,{\ep_{\ba\mu\ba\nu}}^{\ba\rho}{\G}_{+\ba\rho}
\eea and ${ S}_{+\A}={ S}_{+\A}P_+$.
The complex conjugated superalgebra
${\overline{OSp(1;2|C)}}=({ S}_{-\B}={S}^\dagger_{+\B},\bJ_{\ba\mu}^\dagger )$
has the following fermionic sector
\bea
\{{S }_{-\A},{ S}_{-\B}\}&=&-4i\,(C\G_-^{\ba\rho})_{\A\B}\,\bJ^\dagger_{\ba\rho},\qquad
\left[\bJ^\dagger_{\ba\mu},{ S}_{-\A}\right]=-\frac12\,
{{\bf S}_{-\B}\,({\Gamma_{-\ba\mu}})^\B}_\A.
\label{OSp12Cm}\eea
Using \bref{bJdef} and introducing real supercharges
${ \CS}_\A={S}_{+\A}+{ S}_{-\A}$
 one gets the following real superalgebra describing the realification  $OSp_R(1;2|C)$ of $OSp(1;2|C)$ which extends supersymmetrically the Lorentz algebra \bref{O31} by the relation
\be
\{{{\CS }}_{\A},{{\CS}}_{\B}\}=-\,(C\gamma^{{\mu}{\nu}})_{\A\B}\CJ_{{\mu}{\nu}},
\qquad
\left[\CJ_{{\mu}{\nu}},{\CS}_{\A}\right]=-\frac{i}2(\CS \gamma_{{\mu}{\nu}})_{\A}.
\label{OSp14alg22}\ee
The superalgebra $OSp_R(1; 2|C) \oplus  OSp(1;4)$ before the contraction $R\to \infty$ is described by the real generators $ (\CM_{\8\mu\8\nu},\CQ_\A,\CJ_{\mu\nu},\CS_\A)$ and does not contain any internal symmetry generators. Using
 the rescalings \bref{relgenSM3Rplus-1} of $O(3,1)\oplus O(3,2)$ generators and new rescaled supercharges
\be
{\bQ}_{\alpha}={R^{-1/2}}({\CQ}_{\alpha} +{\CS}_{\alpha}),\quad
{\bSig}_{\alpha}={R^{-3/2}}{\CS}_{\alpha},
\label{comgeb}
\ee
we describe the real superalgebra $OSp_R(1; 2|C)\oplus OSp(1|4)$. The bosonic part is given by the relations
\bref{BMaxalgeb02} and the part of superalgebra which contains the supercharges $\bQ_\A$ and $\bSig_\A$ is
\bea
\left[P_\mu,{\bQ}_\A\right]&=&\frac{i}2\,(\bSig\gamma_{\mu})_\A
-\frac{i}{2R}\,(\bQ\gamma_{\mu})_\A,\qquad
\left[P_\mu,{\bSig}_\A\right]=0,
\label{kalgebraPQ2}
\nn\\
\left[Z_{\mu\nu},{\bQ}_\A\right]&=&-\frac{i}2\frac1{R M^2}\,(\bSig\gamma_{\mu\nu})_\A,\qquad
\left[Z_{\mu\nu},{\bSig}_\A\right]=  -\frac{i}2\,\frac1{R^{2}M^2}\,(\bSig\gamma_{\mu\nu})_\A,
\nn\\
\left[M_{\rho\s},\bQ_{\A}\right]&=&-\frac{i}2(\bQ \gamma_{\rho\s})_{\A},\qquad
\left[M_{\rho\s},\bSig_{\A}\right]=-\frac{i}2(\bSig \gamma_{\rho\s})_{\A}\,,
\\
\{{\bQ}_\A,{\bQ}_\B\}&=&-\frac{1}R\,(C\gamma^{\mu\nu})_{\A\B}\,M_{\mu\nu}\,+
2\,(C\gamma^{\mu})_{\A\B}P_\mu,
\nn\\
\{{\bQ}_\A,{\bf\Sigma}_{\B}\}&=&-{M^2}\,(C\gamma^{\mu\nu})_{\A\B}Z_{\mu\nu},
\qquad
\{{\bSig}_\A,{\bSig}_\B\}=-\,\frac{M^2}R\,(C\gamma^{\mu\nu})_{\A\B}Z_{\mu\nu}.
\label{kalgebra2}
\eea
In the limit $R\to\infty$ the superalgebra
$OSp_R(1; 2|C) \oplus  OSp(1;4)$ is contracted into the minimal simple Maxwell superalgebra (see \cite{Bonanos:2009wy}
 and (\ref{Maxsuperalgebra-1}-\ref{Maxsuperalgebra}))
with $B=0$ and removed generator $B_C$.
It should be added that  if we choose $\A=0,\B=1$ the similar contraction formulae were considered recently in \cite{biwroja10}.


\subsection{Contraction of $ OSp_R(2; 2|C)\oplus Sp(4)\quad (k=0$ case)}

For completeness we present the exotic version of Maxwell superalgebra
obtained by the contraction of \bref{wroja1.3} for $k=0$.
The bosonic algebra $Sp(4)=O(3,2)$ is given by relation \bref{OSp04}
and $OSp(2;2|C)$  is the complex superalgebra with bosonic subalgebra
$Sp(2|C)\oplus O(2|C)$ where $O_R(2|C)= O(2)\oplus O(1,1)$.
Fermionic sector of $ OSp_R(2; 2|C)$ is described by the complex $O(2)$ doublet of supercharges ${ S}^i_{+\A}$ as follows
\bea
\{{ S}^i_{+\A},{ S}^j_{+\B}\}&=&4i\,\D^{ij}(C\G_+^{\ba\rho})_{\A\B}\,\bJ_{\ba\rho}+2\,(C P_+)_{\A\B}\,\ep^{ij} \bT_+,
\nn\\
\left[\bJ_{\ba\mu},{ S}^i_{+\A}\right]&=&\frac12\,
{{ S}^i_{+\B}\,({\Gamma_{+\ba\mu}})^\B}_\A,
\qquad
\left[\bT_+,{ S}^i_{+\A}\right]=i\,\ep^{ij}\,{ S}^{j}_{+\A},
\label{OSpCp}\eea
where $\bT_+$ is the complex $O(2|C)$ generator.
The fermionic sector of complex conjugate generators  $(\bJ^\dagger, { S}^{i\dagger}_+, \bT^{\dagger}_+)=
(\bJ^\dagger, { S}^{i}_-, \bT_-)$ of $\,{\overline{OSp(2; 2|C)}}$ satisfy the conjugate relations
\bea
\{{S}^i_{-\A},{ S}^j_{-\B}\}&=&-4i\,\D^{ij}(C\G_-^{\ba\rho})_{\A\B}\,\bJ^\dagger_{\ba\rho}+2\,(C\,P_-)_{\A\B}\,\ep^{ij}\, \bT_-,
\nn\\
\left[\bJ^\dagger_{\ba\mu},{ S}^i_{-\A}\right]&=&-\frac12\,
{{ S}^i_{-\B}\,({\Gamma_{-\ba\mu}})^\B}_\A,\qquad
\left[ \bT_-,{ S}^i_{-\A}\right]=i\,\ep^{ij}\,{ S}^{j}_{-\A}.
\label{OSpCm}\eea
The generators $(\bJ,{ S}^i_+, \bT_+)$ and
$(\bJ^\dagger, { S}^i_-, \bT_-)$ are commuting. The realification of $OSp(2;2|C)$ is the real superalgebra  $OSp_R(2;2|C)$ extending the Lorentz algebra \bref{O31} as follows
\bea
\{{\CS}_{\A}^i,{\CS}_{\B}^j\}&=&-\,\D^{ij}\,(C\gamma^{{\mu}{\nu}})_{\A\B}
\CJ_{{\mu}{\nu}}+\ep^{ij}\,(C(\CT_0+\gamma_5\CT_5))_{\A\B},
\nn\\
\left[\CJ_{{\mu}{\nu}},{ S}_{\A}^i\right]&=&-\frac{i}2(S^i \gamma_{{\mu}{\nu}})_{\A},
\quad
\left[ \CT_0,{\CS}_{\A}^i\right]={i}\ep^{ij}\,{\CS}^j_{\A},\quad
\left[ \CT_5,{\CS}_{\A}^i\right]={i}\ep^{ij}\,({\CS}^i\gamma_5)_{\A},
\label{OSp22C7}\eea
where $
{\CS}_\A^i={S}_{+\A}^i+{S}_{-\A}^i$ and $ \bT_+=\CT_0+i\,\CT_5.$
Let us introduce the rescaled generators by the redefinitions
\bref{relgenSM3Rplus-1} and
\bea
{\bf S}_\A^i&=&\frac1{R} S_\A^i,
\label{geredefba}\\
\bT_0&=&\frac1{R^{c_0}} \CT_0,\qquad \bT_5=\frac1{R^{c_5}} \CT_5.
\label{geredefb}
\eea
In terms of the generators defined by \bref{relgenSM3Rplus-1} and
(\ref{geredefba}-\ref{geredefb})
the superalgebra  $OSp_R(2;2|C)\oplus Sp(4)$ is described by the formulae \bref{BMaxalgeb02} supplemented by the following relations;
\bea
\{{\bS^{i}}_\A,{\bS^{j}}_\B\}
&=&-\,\D^{ij}\,M^2\,(C\gamma^{\mu\nu})_{\A\B}Z_{\mu\nu}
-\,\frac{\A}{R^2}\D^{ij}\,(C\gamma^{\mu\nu})_{\A\B}M_{\mu\nu}
\nn\\&&+\ep^{ij}(\frac1{R^{2-c_0}}C_{\A\B}\bT_0-\frac1{R^{2-c_5}}\,(C\gam_5)_{\A\B}\,\bT_5),\label{eq3.281}\\
\left[M_{\mu\nu},\bS^{i}_\A\right]&=&-\frac{i}{2}\,
(\bS^{i}\gam_{\mu\nu})_\A,\qquad
\left[Z_{\mu\nu},\bS^{i}_\A\right]=-\frac{i\B}{2M^2R^2}\,
(\bS^{i}\gam_{\mu\nu})_\A,\nn\\
\left[\bT_0,\bS^{i}_\A\right]&=&\frac{i}{R^{c_0}}
\ep^{ij} \bS^{j}_\A,\qquad
\left[\bT_5,\bS^{i}_\A\right]=\frac{i}{R^{c_5}}
\ep^{ij}  (\bS^{j}\gamma_5)_{\A}.\label{eq3.53}\eea
If $c_0=c_5 = 2$ one gets in the contraction limit $R\to\infty$ the N=1 exotic version of Maxwell  superalgebra with the basic
  superalgebra relation having the form \bref{wroja1.5} obtained from \bref{eq3.281}, and includes nonvanishing
 central charges $\bT_0$ and $\bT_5$. By choosing
 $2> c_0, c_5 >0$
these  generators are decoupled in the   contraction limit, and for
 $c_0=c_5=0$ we obtain $\bT_0$ and $\bT_5$ as describing the Abelian $O(2)$
symmetry generators.

\vs
\section{$N$-extended Maxwell superalgebras}

In previous section we did see that the fermionic sector (${\bf Q}_\alpha, {\bf \Sigma}_\alpha$) of simple Maxwell algebra can be obtained by the contractions from the pair of $OSp(2;4)$ supercharges
${ Q}^{i}_{\alpha}$ ($i=1,2$) or from the pair of supercharges
${ Q}_{\alpha} \in OSp(1;4)$ and ${ S}_\alpha \in OSp_R(1,2|C)$. It is interesting to see what variety of contractions can be obtained if $N>1$, i.e. for the superalgebras \bref{wroja1.3} with $2N$ Weyl supercharges.

The extension to $N>1$ of two types of contraction presented in Sect.2 is based on the observation that the standard Poincar\'{e} supercharges
satisfying relation \bref{wroja1.4} and  belonging to Maxwell superalgebra
  can be obtained by contraction in two ways

 \begin{description}
 \item[i)] from $n$ pairs of $OSp(2n;4)$ supercharges (see \bref{relgenSM3Rplus}) one gets ${\bf Q}^{i}_{\alpha}$ with $i=1\ldots n$,

 \item[ii)] from $m$ pairs of supercharges of the superalgebra
  ${OSp_R(m;2|C)}\oplus OSp(m;4)$ (see \bref{comgeb}) providing $m$
Poincar\'e supercharges ${\bf Q}^i_{\alpha},\;(i=1\ldots m)$.
    \end{description}

 In general case we consider the contractions of ${OSp_R(r;2|C)}\oplus OSp(k;4)$ with $k+r=2N$.
 The first type {\bf  i)} of contraction describes all  Poincar\'{e} superalgebra generators only
   if $k=2N$ - one gets $n=N$. The second way {\bf  ii)}  of contracting  describes the whole Poincar\'{e} supercharges sector only if $k=N$  - again we obtain $N$ Poincar\'{e} supercharges.
  If $k\neq 0, N,2N$ one should apply for supercharges both types of contractions {\bf  i)} and {\bf  ii)}.

We shall distinguish the following two separate cases;
  \begin{description}
  \item[a)] $2N \geq k \geq N\geq r$ where $k+r=2N$.
  \\
  The supercharges in the superalgebra \bref{wroja1.3} can be described by
  $({ Q}^{i}_{\alpha}, { Q}^{i'}_{\alpha},  S^{i}_{\alpha}
, (i=1,...,r,\, i'=r+1,...,k).$
  From the $r$ pairs $({ Q}^{i}_{\alpha}, { S}^{i}_{\alpha})\;$
 one obtains $r$ Poincar\'{e} supercharges via the second mechanism
{\bf ii)};
   remaining even number $k-r=2(k-N)$ of supercharges
   ${ Q}^{i'}$ produce $k-N$ Poincar\'{e} supercharges as in {\bf i)}.
Concluding, we obtain in the contraction limit $r+(k-N)=N$ Poincar\'{e} supercharges satisfying the relation \bref{wroja1.4}.
       \item[b)]  $2N \geq r \geq N\geq k$ where  $k+r=2N$.
   \\
   We get the supercharges (${ Q}^{i}_{\alpha}, { S}^{i}_{\alpha},
  {S}^{i''}_{\alpha}),\; (i=1,...,k,\, i''=k+1,...,r)$
    and we can obtain
from the $k$ pairs $({ Q}^{i}_{\alpha}, { S}^{i}_{\alpha}), (i=1,...,k),$
only $k$ Poincar\'{e} supercharges by the second mechanism (see {\bf ii)}).
The remaining $r-k=2(N-k)$ supercharges lead after contraction to the
fermionic
sector with exotic supercharges, satisfying the relation \bref{wroja1.5}.
\end{description}
The case $k=N$ is described equally well by both cases {\bf a)} and/or {\bf b)}.

\subsection{The superalgebras $OSp(k;4)$ and ${OSp_R(r;2|C)}$ }

The $OSp(k,4)$  superalgebra extends supersymmetrically
the $O(3,2)$ algebra   \bref{OSp04} as follows
\bea
\{{\CQ}_{\A}^{ i},{\CQ}_{\B}^{ j}\}&=&-\,\D^{ij}\,(C\Gam^{{\8\mu}{\8\nu}})_{\A\B}
\CM_{{\8\mu}{\8\nu}}+\,2\,(C)_{\A\B}\CB^{ij},
\nn\\
\left[\CM_{{\8\mu}{\8\nu}},\CQ_{\A}^{ i}\right]&=&-\frac{i}2(\CQ^i \Gam_{{\8\mu}{\8\nu}})_{\A},\qquad
\left[\CB^{ij},{\CQ}_{\A}^{\l}\right]=-i\,\D^{\l[j}\,\CQ_{\A}^{i]},
\nn\\
\left[\CB^{ij},{\CB}^{\l m}\right]&=&-i\,\D^{\l[j}\,\CB^{i]m}
+i\,\D^{m[j}\,\CB^{i]\l}.
\label{4OSp24}\eea
The supercharges ${\CQ}_{\A}^{i},\,(\A=1,2,3,4,\,i=1,2,...,k)$ are real $SO(3,2)$ spinors and $\CB^{ij}=-\CB^{ji}\, (i,j=1,2,...,k)$ are $O(k)$ generators.
After 4+1 decomposition of the Lorentz indices $\8\mu=(\mu,4)$ the superalgebra \bref{4OSp24}  can be written as $D=4$ super-AdS algebra, where $\CP_\mu=\CM_{\mu 4}$
\bea
\left[\CM_{{\mu}{\nu}},\CM_{\rho\s}\right]&=&
-i\,\h_{{\rho}[\nu}\CM_{{\mu]}\s}+i\,
\h_{{\s}[\nu}\CM_{{\mu]}\rho]},
\nn\\
\left[\CP_{{\mu}},\CM_{\rho\s}\right]&=&
-i\,\h_{{\mu}[\rho}\CP_{\s]},\qquad
\left[\CP_{{\mu}},\CP_{\nu}\right]=
-i\,\CM_{\mu\nu},
\nn\\
\{{\CQ}_{\A}^{ i},{\CQ}_{\B}^{ j}\}&=&2\,\D^{ij}\,(C\gam^{\mu})_{\A\B}
\CP_{{\mu}} \,-\D^{ij}\,(C\gam^{{\mu}{\nu}})_{\A\B}\CM_{{\mu}{\nu}}\,
+2\,(C)_{\A\B}\CB^{ij},
\nn\\
\left[\CM_{{\mu}{\nu}},\CQ_{\A}^{ i}\right]&=&-\frac{i}2(\CQ^i \gam_{{\mu}{\nu}})_{\A},\qquad
\left[\CP_{{\mu}},\CQ_{\A}^{ i}\right]=-\frac{i}2(\CQ^i \gam_{{\mu}})_{\A},
\nn\\
\left[\CB^{ij},{\CQ}_{\A}^{\l}\right]&=&-i\,\D^{\l[j}\,\CQ_{\A}^{ i]},\qquad
\left[\CB^{ij},{\CB}^{\l m}\right]=-i\,\D^{\l[j}\,\CB^{i]m}
+i\,\D^{m[j}\,\CB^{i]\l}.
\label{OSp242}\eea

The $ OSp(r; 2|C)$ superalgebra has the following form
\bea
\left[\bJ_{\ba\mu},\bJ_{\ba\nu}\right]&=&{i}\,{\ep_{\ba\mu\ba\nu}}^{\ba\rho}
\bJ_{\ba\rho},\qquad
\left[{\bf T}_+^{ij},{\bf T}_+^{\l m}\right]=-i\,(\D^{\l[j}\,{\bf T}_+^{i]m}-
\D^{m[j}\,{\bf T}_+^{i]\l}),
\nn\\
\{{S}^i_{+\A},{S}^j_{+\B}\}&=& 2i \left(2\,\D^{ij}(C\G_+^{\ba\rho})_{\A\B}\,\bJ_{\ba\rho}-i
\,(C P_+)_{\A\B}\,{\bf T}_+^{ij}\right),
\nn\\
\left[\bJ_{\ba\mu},{S}^i_{+\A}\right]&=&\frac12\,
{S^i_{+\B}\,({\Gamma_{+\ba\mu}})^\B}_\A,\qquad
\left[{\bf T}_+^{ij},{S}^{\l}_{+\A}\right]=-i\,\D^{{\l}[j}\,S^{i]}_{+\A},
\label{OSpNCp}\eea
where $\bJ_{\ba\mu},(\ba\mu=0,1,2)$, ${\bf T}_+^{ij}, (i=1,...,r)$ and
$S^i_{+\A}$ are complex $SO(2,1)$, $O(r)$ and supersymmetry
generators respectively.
The complex conjugate generators
$(\bJ, S^{i}_+,{\bf T}_+)^{\dagger}=
(\bJ^\dagger, S^{i}_-,{\bf T}_-)$ describing the superalgebra
${\overline{OSp({r},2|C)}}$ satisfying the conjugate relations (for $r=2$
see also \bref{OSpCm}).
One obtain finally the following real $OSp_R(r;2|C) $ superalgebra
\bea
\left[\CJ_{{\mu}{\nu}},\CJ_{\rho\s}\right]&=&-i\,
\h_{{\rho}[\nu}\CJ_{{\mu]}\s}+i\,\h_{{\s}[\nu}\CJ_{{\mu]}\rho},
\qquad
\left[\CBA^{ij},\CBA^{k{\ell}}\right]=-{i}(\D^{k[j}\,\CBA^{i]{\ell}}-
\D^{\l[j}\,\CBA^{i]k}),
\nn\\
\left[\CBA^{ij},\CBB^{k{\ell}}\right]&=&-{i}(\D^{k[j}\,\CBB^{i]{\ell}}-
\D^{{\ell}[j}\,\CBB^{i]k}),\qquad
\left[\CBB^{ij},\CBB^{k{\ell}}\right]={i}(\D^{k[j}\,\CBA^{i]{\ell}}-
\D^{\l[j}\,\CBA^{i]k}),\nn\\
\{{\CS}_{\A}^i,{\CS}_{\B}^j\}&=&-\,\D^{ij}\,(C\Gam^{{\mu}{\nu}})_{\A\B}\CJ_{{\mu}{\nu}}+\,(C({\CBA}^{ij}-\gamma_5\CBB^{ij}))_{\A\B},\label{OSp22C8}\\
\left[\CJ_{{\mu}{\nu}},\CS_{\A}^i\right]&=&-\frac{i}2(\CS^i \gam_{{\mu}{\nu}})_{\A},\quad
\left[\CBA^{ij},\CS_{\A}^k\right]=-{i}\D^{k[j}\,\CS^{i]}_{\A},\quad
\left[\CBB^{ij},\CS_{\A}^k\right]=-{i}\D^{k[j}\,(\CS^{i]}\gamma_5)_{\A},
\nn\\ \label{OSp2CRalg}\eea
where the real generators $\CJ_{\mu\nu}, \CS^i_\A, \CBA^{ij}, \CBB^{ij}$ describe respectively
$SO(3,1)$, supersymmetry and the pair of internal $SO(r)$ generators that are related to the complex ones by the formula \bref{bJdef} and the relations
$
{S}_{+\A}^i+{S}_{-\A}^i=\CS_\A^i,\, $ $
{\bf T}_+^{ij}=\frac12(\CBA^{ij}+i\,\CBB^{ij}).$


\subsection{Contraction of $OSp_R(r;2|C)\oplus OSp(k;4),\; (k\geq r\geq 0)$}

In this subsection we consider the case {\bf a)} $(k\geq r)$.
In order to obtain after contraction $R\to\infty$ the Maxwell algebra as bosonic subalgebra we rescale the generators $(\CM_{\mu\nu},\CP_\mu,\CJ_{\mu\nu})\in O(3,2)\oplus O(3,1)$ in accordance with the formulae \bref{relgenSM3Rplus-1} .
In the contraction limit $(R\to\infty)$ we obtain the bosonic Maxwell
algebra \bref{wroja1b} for any value of $(\A,\B=1-\A)$.

The generators of  the internal symmetries $ O(r|C)\oplus O(k), \;(k\geq r)$
are split into three families,
\bea
 1) &&(\CB^{ij},\CT_0^{ij},\CT_5^{ij})\in O(r)\oplus O_R(r|C), \quad (i,j=1,...,r),\; \nn\\
2) && \CB^{i'j'} \in O(k-r)=O(2(k-N)), \quad (i',j'=r+1,...,k),\quad
\label{intgene}\\
 3) && \CB^{ij'} \in O(k)/(O(r)\oplus O(k-r)), \quad (i=1,...,r,\, j'=r+1,...,
k).\nn\eea
In the first group of the generators \bref{intgene} the diagonal subalgebra $O(r)$ remain unscaled
\be \bB_D^{ij}=\CB^{ij}+\CT_0^{ij} \label{eq3.9}\ee
while the remaining ones are rescaled as \footnote{
In general case one can consider the rescaling
$\bT_0^{ij}=\frac1R((1-\A')\,\CT_0^{ij}-\A'\,\CB^{ij})$. It appears however that the contraction limit $R\to\infty$ will not depend on $\A'$, so we choose for simplicity $\A'=0$.}
\be
\bT_0^{ij}=\frac1R\,\CT_0^{ij},\qquad
 \bT_5^{ij}=\frac1R\,\CT_5^{ij}.
\label{BTrescaleb} \ee
The generators $\CB^{i'j'} =\CB^{i'j'}_++\CB^{i'j'}_-\in O(k-r)$ we
decompose as follows
\be 
\CB_-^{i'j'}\in U(k-N),\quad
\CB_+^{i'j'}\in O(2(k-N))/U(k-N),
\label{defBpm}\ee
where $2(k-r)\times 2(k-r)$ matrix of generators $\CB=(\CB^{i'j'})$ is decomposed as follows
\be
\CB=
\begin{pmatrix}A_0 & S \\ -S & A_0 \end{pmatrix}+\begin{pmatrix}
A_3 & A_1 \\ A_1 & -A_3 \end{pmatrix}=\CB_-+\CB_+,
\ee
where the matrices $S$ and $A_\l\, (\l=1,2,3)$ satisfy $S^T=S$ and $A_\l^T=-A_\l$.
If we introduce the anti-symmetric matrix $\W=\begin{pmatrix}
0 & 1_{k-N} \\ -1_{k-N} & 0 \end{pmatrix}$, the matrices $\CB_\mp$ satisfy the
relations $ \W\, \CB_\pm \pm \CB_\pm\,\W=0. $
The generators $\CB^{i'j'}_\mp $ we rescale as follows
\be
\bB^{i'j'}_-=\,\CB^{i'j'}_-,\qquad
\bB^{i'j'}_+=\frac1R\,\CB^{i'j'}_+.
\label{eq3.14}\ee
The remaining internal symmetry generators  $\CB^{ij'}$ which occur in the anti-commutators of $\{\bQ^i,\bQ^{j'}\}$ and $\{\bQ^i,\bSig^{j'}\}$ are rescaled by
\be
\bB^{ij'}=\frac1R\,\CB^{ij'}.
\label{eq3.15}\ee

If we perform the contraction $R\to\infty$ of the internal symmetry generators listed in \bref{intgene}  we obtain the pair of non-Abelian Lie algebras $O(r)$ (generators $\bB_D^{ij})$ and $U(N-r)$ (generators $\bB_-^{i'j'})$; remaining contracted generators $\bT_0^{ij}, \bT_5^{ij}, \bB_+^{i'j'}, \bB^{ij'}$ are becoming Abelian.

The $k$ supercharges $\CQ^i,(i=1,...,k)$ are split into $r$  supercharges $\CQ^i,(i=1,...,r)$ and remaining  $k-r$  supercharges $\CQ^{i'},(i'=r+1,...,k)$.
The first ones are combined with  $r$  supercharges $\CS^i,(i=1,...,r)$
to define
\be
\bQ^i_\A=\frac1{R^{1/2}}\,(\CQ^i_\A +\CS^i_\A ) ,\qquad
\bSig^i_\A =\frac1{R^{3/2}} \,\CS^i_\A.
\label{eq3.16}\ee
The remaining $\CQ^{i'},(i'=r+1,...,k)$ are used in order to define
\be
{\bQ^{i'}_\A }= \frac{1}{R^{1/2}}\,\CQ^{j'}_\B \,{{\Pi^+}^{j'i'\B}}_{\A},
\qquad
{\bSig^{i'}_\A }=\frac{1}{R^{3/2}}\,\CQ^{j'}_{\B}\,{{\Pi^-}^{j'i'\B}}_{\A} ,
\label{eq3.17}\ee
where $\Pi^\pm$ are the projection operators ( we recall that $\gamma_5^2=\W^2=-1$),
\be
{{\Pi^\pm}^{j'i'\B}}_{\A}=\frac12({{\D}^{j'i'}}{\D^\B}_{\A}\pm{{\gamma_5}^\B}_\A
{{\W}^{j'i'}}),\qquad \Pi^\pm \Pi^\pm=\Pi^\pm,\quad \Pi^\pm \Pi^\mp=0.
\ee
Supercharges $\bQ^{i'},\bSig^{i'}, (i'=r+1,...,k)$ satisfy
the identities
\be \bQ^{i'}=\bQ^{j'}\,\Pi^{+j'i'},\qquad \bSig^{i'}=\bSig^{j'}\,\Pi^{-j'i'},
\ee
but because $k-r=2(k-N)$ the number of independent supercharge components is reduced to $k-N$ for both  $\bQ^{i'}$ and $\bSig^{i'}$.

One can write down the superalgebra $OSp_R(r;2|C)\oplus OSp(k;4)$
in terms of rescaled generators $(P_\mu,M_{\mu\nu},Z_{\mu\nu},\bQ^i,\bQ^{i'},\bSig^i,\bSig^{i'},\bB_D^{ij},\bT_0^{ij},\bT_5^{ij},\bB_-^{i'j'},\bB_+^{i'j'},\bB^{ij'})$, (see \bref{relgenSM3Rplus-1}, (\ref{eq3.9}-\ref{BTrescaleb}), (\ref{eq3.14}-\ref{eq3.17})). If we perform the contraction limit $R\to\infty$ we obtain the following relations describing $N$-extended $(N>1)$ Maxwell superalgebras,
\bea
\left[P_\mu,P_{\nu}\right]&=&i\,M^2\,Z_{\mu\nu},\qquad
\left[M_{\mu\nu},M_{\rho\s}\right]=-i\,\h_{\rho[\nu}M_{\mu] \s}+i\,
\h_{\s[\nu}M_{\mu]\rho},\nn\\
\left[P_\mu,M_{\rho\s}\right]&=&-i\,\h_{\mu[\rho}P_{\s]},\qquad
\left[M_{\mu\nu},Z_{\rho\s}\right]=-i\,\h_{\rho[\nu}Z_{\mu] \s}+i\,
\h_{\s[\nu}Z_{\mu]\rho},
\eea
\bea
\left[P_\mu,\bQ^i\right]&=&\frac{i}{2}\,\bSig^i\,\gam_\mu,\quad
\left[M_{\mu\nu},\bQ^i\right]=-\frac{i}2\,\bQ^i\gam_{\mu\nu},\quad
\left[M_{\mu\nu},\bSig^i\right]=-\frac{i}2\,\bSig^i\gam_{\mu\nu},
\nn\\
\left[P_\mu,\bQ^{i'}\right]&=&-\frac{i}{2}\,\bSig^{i'}\,\gam_\mu,\quad
\left[M_{\mu\nu},\bQ^{i'}\right]=-\frac{i}2\,\bQ^{i'}\gam_{\mu\nu},\quad
\left[M_{\mu\nu},\bSig^{i'}\right]=-\frac{i}2\,\bSig^{i'}\gam_{\mu\nu},
\nn\\
\left[Z_{\mu\nu},\bQ^{i}\right]&=&
\left[Z_{\mu\nu},\bQ^{i'}\right]=
\left[Z_{\mu\nu},\bSig^{i}\right]=\left[Z_{\mu\nu},\bSig^{i'}\right]=0, \eea
\bea
\{{\bQ^i}_\A,{\bQ^j}_\B\}&=&
2\,\D^{ij}\,(C\gamma^{\mu})_{\A\B}P_\mu
-\,C_{\A\B}\,\bT_0^{ij}-\,(C\gam_5)_{\A\B}\,\bT_5^{ij},\quad
\nn\\
\{{\bQ^i}_\A,{\bSig^j}_\B\}&=&
-\,{\D^{ij}}\,(C\gamma^{\mu\nu})_{\A\B}M^2\,Z_{\mu\nu},
\nn\\
\{{\bQ^i}_\A,{\bQ^{j'}}_\B\}&=&{2}\,(C\bB^{i,\l'}\Pi^{+\l'j'})_{\A\B}.
\nn\\
\{{\bQ^{i'}}_\A,{\bQ^{j'}}_\B\}&=&2(C\gam^\mu\Pi^+)_{\A\B}^{i'j'}\,P_\mu
+\,{2}\,(C\bB_+\Pi^+)^{i',j'}_{\A\B},
\nn\\
\{{\bQ^{i'}}_\A,{\bSig^{j'}}_\B\}&=&(C\gam^{\mu\nu}\Pi^-)_{\A\B}^{i'j'}\,M^2\,Z_{{\mu}{\nu}}.\label{QQalg}\eea
The internal symmetry sector containing the generators $\bB_D^{ij}\in O(k)$
and
$\bB_-^{i'j'}\in U(N-k)$ is described by the following non-vanishing commutators,
\bea
\left[\bB_D^{ij},\bQ^\ell_\A\right]&=&-i\D^{\ell[j} \bQ^{i]}_{\A},\qquad
\left[\bB_D^{ij},\bSig^\ell_\A\right]=-i\D^{\ell[j} \bSig^{i]}_{\A},
\nn\\
\left[\bB_D^{ij},\bB_D^{\ell m}\right]&=&-i(\D^{\ell[j} \bB_D^{i]m}-
\D^{m[j} \bB_D^{i]\ell}),\qquad
\left[\bB_D^{ij},{\bT_0}^{\ell m}\right]=-i(\D^{\ell[j} \bT_0^{i]m}-
\D^{m[j} \bT_0^{i]\ell}),
\nn\\
\left[\bB_D^{ij},\bT_5^{\ell m}\right]&=&-i(\D^{\ell[j} \bT_5^{i],m}-
\D^{m[j} \bT_5^{i]\ell})
,\qquad
\left[\bB_D^{ij},\bB^{\ell m'}\right]=-i\D^{\ell[j} \bB^{i]m'},
\eea
\bea
\left[\bT_0^{ij},\bQ^\ell_\A\right]&=&-i\D^{\ell[j} \bSig^{i]}_{\A},
\quad
\left[\bT_5^{ij},\bQ^\ell_\A\right]=-i\D^{\ell[j} \bSig^{i]}_{\A}\gamma_5,
\quad
\left[\bB^{ij'},\bQ^\ell_\A\right]={i}\D^{i\ell} \bSig^{j'}_\A,
\nn\\
\left[\bB^{ij'},\bQ^{\ell'}_\A\right]&=& {i}( \bSig^{i}\Pi^{+j'\l'})_\A,
\quad
\left[\bB_-^{i'j'},\bQ^{\ell'}_\A\right]=
-i\bQ^{[i'}\,\Pi^{+j']\l'}, \quad
\left[\bB_+^{i'j'},\bQ^{\ell'}_\A\right]
=-i\bSig^{[i'}\,\Pi^{+j']\l'}, \nn\\
\left[\bB_-^{i'j'},\bSig^{\ell'}_\A\right]&=&-i\bSig^{[i'}\,\Pi^{-j']\l'},
\quad
\left[\bB_-^{i'j'},\bB^{\ell m'}\right]=
{i}\,\rho^{ij,mn}_-\,\bB^{\l n'}.\eea


We see from \bref{QQalg} that all fermionic generators $\bQ^i_\A$ and $\bQ^{i'}_\A$ are the Poincar\'e supercharges.

\subsection{Contraction of $OSp_R(r;2|C)\oplus OSp(k;4),\; (r\geq k\geq 0)$}

 The generators of  the internal symmetries $ O(r|C)\oplus O(k), \;(k\leq r)$
are split into the following three families,
\bea
 1) &&(\CB^{ij},\CT_0^{ij},\CT_5^{ij})\in O(k)\oplus O_R(k|C), \quad (i,j=1,...,k),\; \nn \\
2) && (\CT_0^{i''j''}, \CT_5^{i''j''}) \in O_R(r-k|C)=O_R(2(r-N)|C), \quad (i'',j''=k+1,...,r),\quad  \nn \\
 3) && (\CT_0^{ij''}, \CT_5^{ij''})\in O_R(r|C)/(O_R(k|C)\oplus O_R(r-k|C)),
\quad (i=1,...,k,\, j''=k+1,...,r).\nn \eea
In the first group we rescale the generators $\bB^{ij},\bT_0^{ij},\bT_5^{ij}$
as in previous subsection (see (\ref{eq3.9}-\ref{BTrescaleb})).
The generators
$ (\CT_0^{i''j''}, \CT_5^{i''j''}) $ and $(\CT_0^{ij''}, \CT_5^{ij''})$ are rescaled as follows
\be
\bT_0^{i''j''}=\frac1{R^2}\CT_0^{i''j''},\quad
\bT_5^{i''j''}=\frac1{R^2}\CT_5^{i''j''},\quad
\bT_0^{ij''}=\frac1{R^{3/2}}\CT_0^{ij''},\quad
\bT_5^{ij''}=\frac1{R^{3/2}}\CT_5^{ij''}.
 \ee
After the contraction limit $R\to\infty$ only the generators $\bB_D^{ij}$ describe the non-Abelian $O(k)$ generators; remaining  generators $ (\bT_0^{ij},
\bT_5^{ij},\bT_0^{ij''}, \bT_5^{ij''},\bT_0^{i''j''}, \bT_5^{i''j''}) $
are becoming Abelian.

The $r$ supercharges $\CS^i,(i=1,...,r)$ are split into $k$  supercharges
$\CS^i,(i=1,...,k)$ and remaining  $r-k$  supercharges $\CS^{i''},(i''=k+1,...,r)$.
The first ones are combined with  $k$  supercharges $\CQ^i,(i=1,...,k)$
to define
\be
\bQ^i_\A=\frac1{R^{1/2}}\,(\CQ^i_\A +\CS^i_\A ) , \qquad
\bSig^i_\A = \frac1{R^{3/2}} \,\CS^i_\A .
\ee
The remaining generators $\CS^{i''},(i''=k+1,...,r)$ are rescaled as
\bea
\bS^{i''}=\frac1{R}\,\CS^{i''}.
\eea
In the contraction limit $R\to\infty$ we obtain besides the Maxwell algebra
 \bref{wroja1b} the following set of algebraic relations
\bea
\left[M_{\mu\nu},\bQ^i\right]&=&-\frac{i}2\,\bQ^i\gam_{\mu\nu},\quad
\left[M_{\mu\nu},\bSig^i\right]=-\frac{i}2\,\bSig^i\gam_{\mu\nu},\quad
\left[M_{\mu\nu},\bS^{i''}\right]=-\frac{i}2\,\bS^{i''}\gam_{\mu\nu},
\nn\\
\left[P_\mu,\bQ^i\right]&=&\frac{i}{2}\,\bSig^i\,\gam_\mu,\quad
\left[\bT_0^{ij},\bQ^\ell_\A\right]=-i\D^{\ell[j} \bSig^{i]}_{\A},\quad
\left[\bT_5^{ij},\bQ^\ell_\A\right]=-i\D^{\ell[j} \bSig^{i]}_{\A}\gam_5.
\eea
\bea
\{{\bQ^i}_\A,{\bQ^j}_\B\}&=&
2\,\D^{ij}\,(C\gamma^{\mu})_{\A\B}P_\mu
-\,C_{\A\B}\,\bT_0^{ij}-\,(C\gam_5)_{\A\B}\,\bT_5^{ij},\quad
\nn\\
\{{\bQ^i}_\A,{\bSig^j}_\B\}&=&
-\,{\D^{ij}}\,(C\gamma^{\mu\nu})_{\A\B}\,M^2\,Z_{\mu\nu},
\nn\\
\{{\bQ^i}_\A,\bS^{j''}_\B\}&=&
(C(\bT_0^{ij''}-\gam_5 \bT_5^{ij''}))_{\A\B},
\nn\\
\{{\bS^{i''}}_\A,{\bS^{j''}}_\B\}
&=&-\,\D^{i''j''}\,(C\gamma^{\mu\nu})_{\A\B}\,M^2\,Z_{\mu\nu}
+(C_{\A\B}\bT_0^{i''j''}-\,(C\gam_5)_{\A\B}\,\bT_5^{i''j''}).
\label{eq3.32}\eea
\bea
\left[\bB_D^{ij},\bQ^\ell_\A\right]&=&-i\D^{\ell[j} \bQ^{i]}_{\A},\quad
\left[\bB_D^{ij},\bSig^\ell_\A\right]=-i\D^{\ell[j} \bSig^{i]}_{\A},
\quad
\left[\bB_D^{ij},\bB_D^{\ell m}\right]=-i(\D^{\ell[j} \bB_D^{i]m}-
\D^{m[j} \bB_D^{i]\ell}),\nn\\
\left[\bB_D^{ij},{\bT_0}^{\ell m}\right]&=&-i(\D^{\ell[j} \bT_0^{i]m}-
\D^{m[j} \bT_0^{i]\ell}),\quad
\left[\bB_D^{ij},\bT_5^{\ell m}\right]=-i(\D^{\ell[j} \bT_5^{i],m}-
\D^{m[j} \bT_5^{i]\ell})
,\qquad
\nn\\
\left[\bB_D^{ij},\bT_0^{\ell m''}\right]&=&-i\D^{\ell[j} \bT_0^{i]m''},
\qquad
\left[\bB_D^{ij},\bT_5^{\ell m''}\right]=-i\D^{\ell[j} \bT_5^{i]m''}.
\eea
We see from relations \bref{eq3.32} that the Maxwell superalgebras derived in this subsection have hybrid structure: the anti-commutators of $k$ Weyl supercharges $\bQ^i$ as in Poincar\'e SUSY algebra
describe the fourmomenta generators $P_\mu$ (see \bref{wroja1.4}), but remaining $r-k$ supercharges $\bS^{i''}$ supersymmetrizing the tensorial generators $Z_{\mu\nu}$ describe the exotic sector of $N$-extended Maxwell superalgebra with the SUSY relations \bref{wroja1.5} .


\section{Discussion}

The aim of this paper is to provide a large class of superalgebraic structures describing $N$-extended Maxwell superalgebra. In Sect.2 we considered simple Maxwell superalgebras $(N=1)$; the extended case $(N>1)$ is described in Sect.3.

For arbitrary $N$ we can considered the contractions of $2N$ different superalgebras described by  \bref{wroja1.3} with $k=1,2,...,2N$. The case $k=0$ as it was argued in introduction and in subsections 2.4 and 3.3 provides only the purely exotic supersymmetrization of Maxwell algebra with all supercharges satisfying the relations \bref{wroja1.5}. If $N\leq k\leq 2N$ we obtain the Maxwell superalgebra with $N$ "standard" SUSY relations \bref{wroja1.4}; if $0\leq k<N$ one obtains a hybrid Maxwell superalgebra, with $k$ standard Poincar\'{e} supercharges and $r-k$ supercharges satisfying exotic SUSY relation  \bref{wroja1.5}.
At present it is not known whether the exotic SUSY relation  \bref{wroja1.5} can provide some dynamical models with physically interesting consequences
but for completeness of our algebraic considerations we included as well the extended Maxwell superalgebras with exotic supercharges.

The contractions considered in this paper can be further applied to the description of various Maxwell supergravity models. For example in order to obtain $N=1$ Maxwell supergravity one should consider firstly the gauge fields on $O(3,1)\oplus OSp(2,4)$ or $OSp_R(1,2|C)\oplus OSp(1,4)$ superalgebras and then introduce the gauge formulations of deformed Maxwell supergravity models.
In the following step one can look for the deformed Maxwell supergravity
actions with finite contraction limits
defined in accordance with the contraction prescription in Sect.2.2 and 2.3. In general case one can consider the gauge theories on superalgebras \bref{wroja1.3} and study by their suitable contractions the possible new extended Maxwell supergravity models.

\vs

{\bf  Acknowledgements }
The authors would like to thank Jose A.~de Azc\'{a}rraga, Joaquim Gomis and Piotr Kosinski for their valuable remarks and discussions. The paper has been supported by Polish NCN grant No.2011/ST2/03354.


\end{document}